\documentclass[aip,apl,reprint,amsmath,amssymb,floatfix]{revtex4-1}

\usepackage[colorlinks,citecolor=blue,linkcolor=blue]{hyperref}
\usepackage{graphicx}% Include figure files
\usepackage{dcolumn}% Align table columns on decimal point
\usepackage{bm}% bold math
\usepackage{txfonts}

\newcommand{\CNN}{Centre de Nanosciences et de Nanotechnologies, CNRS, Univ. Paris-Sud, Universit\'e Paris-Saclay, 91120 Palaiseau, France}
\newcommand{\LMOPSa}{Chaire Photonique, Laboratoire LMOPS, CentraleSup\'elec, Universit\'e Paris-Saclay, 57070 Metz, France}
\newcommand{\LMOPSb}{Laboratoire Mat{\'e}riaux Optiques, Photonique et Syst{\`e}mes, CentraleSup\'elec, Universit\'e de Lorraine, 57070 Metz, France}

\begin{document}

\title{Chaotic dynamics in a macrospin spin-torque nano-oscillator with delayed feedback}

\author{J{\'e}r{\^o}me Williame}
\author{Artur Difini Accioly}
\affiliation{\CNN}
\author{Damien Rontani}
\author{Marc Sciamanna}
\affiliation{\LMOPSa}
\affiliation{\LMOPSb}
\author{Joo-Von Kim}
\email{joo-von.kim@c2n.upsaclay.fr}
\affiliation{\CNN}

\date{\today}% It is always \today, today

\begin{abstract}
A theoretical study of delayed feedback in spin-torque nano-oscillators is presented. A macrospin geometry is considered, where self-sustained oscillations are made possible by spin transfer torques associated with spin currents flowing perpendicular to the film plane. By tuning the delay and amplification of the self-injected signal, we identify dynamical regimes in this system such as chaos, switching between precession modes with complex transients, and oscillator death.  Such delayed feedback schemes open up a new field of exploration for such oscillators, where the complex transient states might find important applications in information processing.
\end{abstract}

\maketitle

%%% Begin text here

%--intro/presentation du systeme--
%\section{Introduction}
Spin-torque nano-oscillators (STNO) are nanoscale electrical oscillators based on ferromagnetic materials that are promising for a number of technological applications, such as microwave sources and field sensors.~\cite{Chen:2016bf, Locatelli2013, Macia2011} They are typically based on magnetoresistive stacks, whereby spin-torques exerted by the flow of spin-polarized currents result in the self-sustained oscillation of the magnetization in the free layer.~\cite{Berkov2008, Li2003, Miltat2002, Kim:2012du} The oscillation state can comprise (quasi-)uniform precession,~\cite{Kiselev:2003hp, Rippard:2004gy} spin wave bullets,~\cite{Slavin:2005es} coupled precession modes in synthetic antiferromagnets~\cite{Firastrau:2013fh, Monteblanco:2013hw} and ferrimagnets,~\cite{Monteblanco:2017kx} gyrating vortices~\cite{Pribiag:2007dk, Pufall:2007jc, Mistral:2008js, Dussaux:2010ef, Locatelli:2011hw} and skyrmions,~\cite{GarciaSanchez:2016cx} and dynamical droplet solitons.~\cite{Mohseni:2013eh}

Delayed feedback in dynamical systems, whereby the output signal of a system is sent back into its input with amplification and delay, can result in a variety of nonlinear behaviors.~\cite{Erneux:2009} One consequence is the possibility of inducing chaotic dynamics in otherwise low-dimensional systems. From a mathematical perspective, delayed feedback extends the original phase space into a theoretically infinite phase space, hence allowing for the observation of chaos of possibly very large dimension. A well-known example is the Mackey-Glass oscillator,~\cite{Mackey:1977dv} which is described by a first-order delay-differential equation and can exhibit a variety of different dynamical states, including limit-cycle and aperiodic states, and complex transients. Nonlinear dynamics from delayed feedback systems has since long been considered for information processing, e.g., secure communications, sensing, lidar, and even machine learning based computing.~\cite{Appeltant:2011jy, Sciamanna:2015ec}

For the STNO, whose dynamics is well-described by a two-dimensional dynamical system~\cite{Kim:2012du}, it is intriguing to inquire whether delayed feedback lead to more complex behavior such as chaos, much like periodic forcing.~\cite{Li:2006gl}  It has been shown that delayed feedback can improve spectral properties such as the emission linewidth.~\cite{Khalsa:2015kn, Tamaru:2016kl, Tsunegi2016} Here, we will present results of a theoretical study on the complex transient response and chaotic behavior in STNOs subject to delayed feedback. We considered a model oscillator system in which the output is generated by changes in the magnetoresistance, which is subsequently fed back as variations in the input drive current. We focus on the macrospin \cite{Xiao:2005cd} oscillator operating near the transition between the in-plane (IPP) and out-of-plane (OPP) precession regimes. By tuning the delay and amplification of the self-injected signal, we identify dynamical regimes in this system such as chaos, IPP/OPP switching with complex transients, and oscillator death.

The macrospin dynamics is described by the Landau-Lifshitz equation with spin torques,~\cite{Slonczewski:1996wo}
\begin{equation}
\frac{d \mathbf{m}}{dt} = -\frac{\gamma_0}{1+\alpha^2} \mathbf{m} \times \mathbf{H}_{\rm eff} + \frac{\gamma_0}{1+\alpha^2} \mathbf{m} \times \left[ \mathbf{m} \times \left(-\alpha \mathbf{H}_{\rm eff} +  J \mathbf{p}  \right) \right],
\label{eq:llg}
\end{equation}
where $\gamma_0 = \mu_0 \gamma$ is the gyromagnetic constant, $\mathbf{m}$ is a unit vector representing the magnetization state, $\mathbf{H}_{\rm eff}$ is the effective field, $\alpha$ is the Gilbert damping constant, $J$ is the applied current density, and $\mathbf{p}$ is orientation of the spin polarization. Note that $J$ is expressed as a magnetic field by using the relation $J = \hbar j / (\mu_0 M_s e d)$, where a density of $j = 10^7$ A/cm$^2$ corresponds to a field of $J = 10$ mT, which is consistent with spin valve nanopillar devices based on  Co/Cu/Co.~\cite{Kiselev:2003hp} In our calculations, we assume a thin film geometry in which $z$ is the direction perpendicular to the film plane with a uniaxial anisotropy and an applied field along the $x$ axis. As such, $\mathbf{H}_\textrm{eff} = (H_0 + H_\textrm{an} m_x)\hat{\mathbf{x}} - H_{d}m_z \hat{\mathbf{z}}$. In what follows, we used $\mu_0 H_0 = 0.1$ T, $\mu_0 H_{\rm an} = 0.05$ T, and $\mu_0 H_{d} = 1.7$ T, which are similar to values considered elsewhere.~\cite{Kiselev:2003hp, Grollier:2006dv} We take $\mathbf{p} = \hat{\mathbf{x}}$ which defines the parallel configuration.

\begin{figure}
\centering\includegraphics[width=7.5cm]{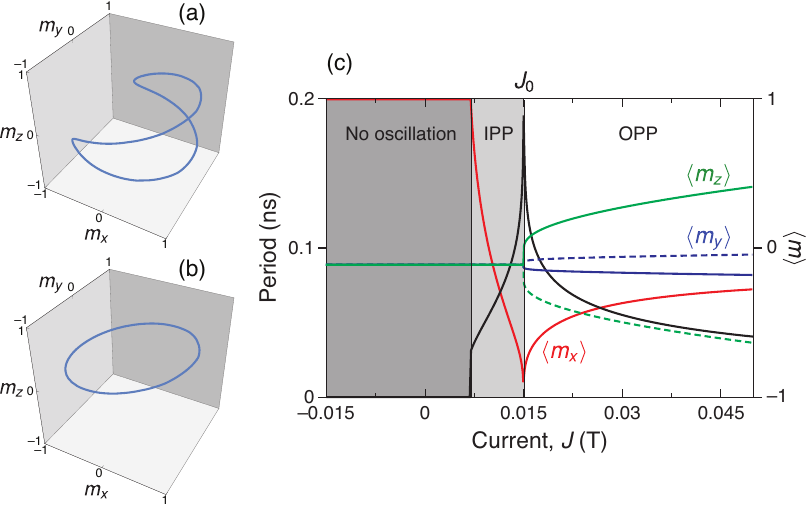}
\caption{Oscillation modes of a macrospin spin-torque nano-oscillator under dc currents. (a) In-plane precession (IPP) under $J=0.01$ T. (b)  Out-of-plane precession (OPP) under $J=0.02$ T. (c) Mean values of the magnetization components and oscillation period as a function of applied current $J$. $J_0$ denotes the operating point.}
\label{fig:states}
\end{figure}

Some possible precession modes are illustrated in Fig.~\ref{fig:states}(a,b).
%%
%% Figure
The onset of self-sustained oscillations first involves precession of the magnetization in the film plane (IPP),~\cite{Xiao:2005cd} where the trajectory has a clamshell shape centered about the $x$ axis [Fig.~\ref{fig:states}(a)]. As the current is increased, the preferred oscillation mode involves out-of-plane precession (OPP), where the axis of precession is the film normal and the orbits are more circular [Fig.~\ref{fig:states}(b)]. There are two degenerate OPP states, i.e., precession about the $+z$ and $-z$ axes, which we denote as (OPP$+$) and (OPP$-$), respectively. The current dependence of the mean values of the three magnetization components and the oscillation period (of the $m_x$ component) are presented in Fig.~\ref{fig:states}(c). We observe a clear current threshold at $J \approx 0.007$ T, below which the magnetization remains static along $x$. Above this threshold in the IPP regime, the average $\langle m_x \rangle$ component (linked to magnetoresistance variations) decreases rapidly as a function of current density, which is also accompanied by a sharp decrease in the oscillation frequency. The average values are $\langle m_y \rangle = \langle m_z \rangle = 0$ in this regime. Above a second threshold, $J \approx 0.015$ T, the system enters the OPP state where all magnetization components have nonzero time averages. The current dependence of $\langle m_x \rangle$ exhibits the opposite behavior compared with the IPP state, where it progressively increases and is accompanied by an increase in the oscillation frequency. The dashed lines in Fig.~\ref{fig:states}(c) indicate the degenerate OPP state.

The output signal of a spin-torque nano-oscillator is typically given by the giant or tunnel magnetoresistance, where the electrical resistance depends on the relative orientation between the free and reference layer magnetizations. It is therefore natural to employ the output current (or voltage) variation as the feedback signal. We assume a time-dependent applied current density of the form
\begin{equation}
J(t)=J_0 \left[ 1+ \Delta j \; m_x (t-\tau) \right],
\label{eq:current}
\end{equation}
where $J_0$ is the injected dc current, $\Delta j$ is the relative feedback amplitude, and $\tau$ is a variable time delay. Since the reference layer polarization $\mathbf{p} = \hat{\mathbf{x}}$, only variations in the $m_x$ component leads to changes in the overall magnetoresistance, which is used as the basis for the feedback signal.

\begin{figure}
\centering\includegraphics[width=7.0cm]{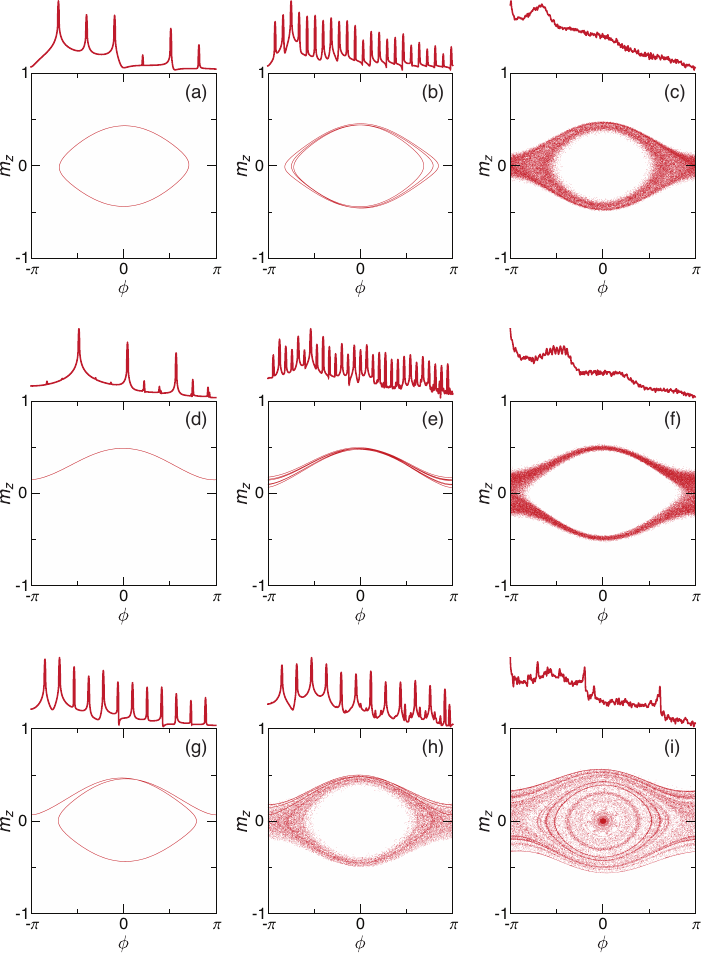}
\caption{Phase portraits of the oscillator dynamics under delayed feedback over 500 ns. For $\Delta j = 1.0$: (a) IPP ($\tau = 0.1$ ns), (b) modulated IPP ($\tau = 0.204$ ns), and (c) chaos ($\tau = 1$ ns). For $\Delta j = -1.0$: (d) OPP ($\tau = 0.135$ ns), (e) modulated OPP ($\tau = 0.15$ ns), and (f) chaos ($\tau = 1$ ns). For $\Delta j = 1.7$: (g) synchronized IPP-OPP ($\tau = 0.0759$ ns), (h) transient chaos ($\tau = 0.174$ ns), and (i) intermittency ($\tau = 13.18$ ns). The inset above each phase portrait shows the power spectrum of the corresponding dynamics, where the horizontal scale represents a range of 50 GHz and the vertical scale represents the power spectral density on a log scale.}
\label{fig:phimz}
\end{figure}

We focus on the feedback dynamics close to the IPP to OPP transition. A constant drive current of $J_0 = 0.015$ T is used, which leads to IPP dynamics but is close to the threshold current for the OPP region. Time delays over several orders of magnitude are considered, which allows different time scales from single precession periods over to longer transients to be probed. Representative trajectories are shown in Fig.~\ref{fig:phimz}.
%%
%% Figure
Because the dynamics of $\mathbf{m}(t)$ is constrained to the unit sphere, it is convenient to examine the trajectories in $(\phi,m_z)$ space, where $\phi = \tan^{-1}(m_y/m_x)$. Besides the IPP and OPP states [Figs.~\ref{fig:phimz}(a) and \ref{fig:phimz}(d), respectively], the delayed feedback can also lead to modulated versions of these states, where distinct orbits for the IPP [Fig.~\ref{fig:phimz}(b], OPP [Fig.~\ref{fig:phimz}(e)], and mixed IPP/OPP [Fig.~\ref{fig:phimz}(g)] can be observed during steady-state oscillation. These steady-state oscillations are characterized by well-defined peaks in the power spectrum. As the time delay is varied, chaotic states appear at positive and negative feedback [Figs.~\ref{fig:phimz}(c) and \ref{fig:phimz}(f), respectively], which are characterized by broad features in the power spectrum across a wide frequency range. We also find evidence of transient chaos [Fig.~\ref{fig:phimz}(h)], where chaotic dynamics is observed over a transient period of a few hundred ns before settling into a modulated OPP trajectory. At long delays, we find cases of intermittency which involve chaotic transitions between long periods of IPP and OPP modes [Fig.~\ref{fig:phimz}(i)]. Oscillator death is also observed under certain conditions (not shown). Schematic illustrations of the power spectra are given as insets above each phase portrait, which are computed over the last 100 ns of the simulation.

\begin{figure*}
\centering\includegraphics[width=12cm]{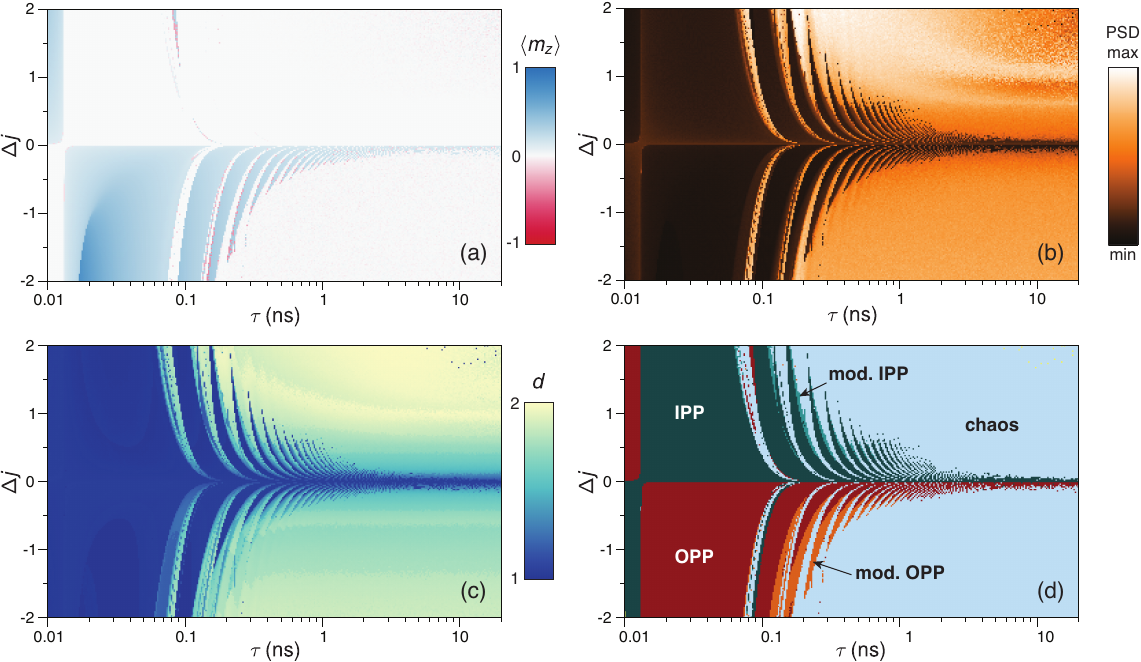}
\caption{Phase diagram of possible dynamics as a function of the feedback amplitude $\Delta j$ and time delay $\tau$. (a) Time averaged $m_z$ component, indicative of OPP. (b) Averaged oscillator power using $m_x$ and $m_z$ components. (c) Dimensionality of trajectories in $(\phi,m_z)$ space. (d) Classification of dynamical regimes identified, where `mod.' denotes modulated states.}
\label{fig:phase}
\end{figure*}

In Fig.~\ref{fig:phase}, we present the full phase diagram of the oscillator behavior as a function of the time delay $\tau$ and feedback amplitude $\Delta j$ with four different representations. Each pixel represents the result of time integrating Eq.~(\ref{eq:llg}) with Eq.~(\ref{eq:current}) over 500 ns.%%
%% Figure
The time-averaged $m_z$ component is shown in Fig.~\ref{fig:phase}(a). With the initial conditions used, the OPP$+$ regimes are primarily visited and distinct bands in their existence can be seen as the delay is varied. A measure of the total oscillator power is given in Fig.~\ref{fig:phase}(b), which is computed by integrating over the power spectral density as shown in the insets of Fig.~\ref{fig:phimz}. Limit cycles lead to low power, as indicated by the black regions, while chaotic dynamics give rise to high powers (orange to white regions). As a complementary measure, we also examined the fractal dimension $d$ of the phase portraits in Fig.~\ref{fig:phimz} with the box-counting method. Limit cycles are represented by lines and have $d=1$, while strongly modulated and chaotic trajectories possess a fractal nature with noninteger $1<d<2$. This analysis is presented in Fig.~\ref{fig:phase}(c), where we can observe distinct bands of steady-state oscillation, with a variety of fractal states that dominate the dynamics at large delays. We note that the fractal dimension does not appear to vary much with the delay at a given value of the feedback amplitude. By combining these measures with the behavior identified without feedback [Fig.~\ref{fig:states}(a)], we construct phase diagram of possible states in Fig.~\ref{fig:phase}(d). IPP states are primarily seen at positive feedback, while OPP states appear for negative feedback. This results from the operating point, where increases in $J_0$ drive the dynamics into the OPP regime, while decreases in the current $J_0$ further stabilize the IPP dynamics. Since $\langle m_x \rangle < 0$ at $J_0$ [Fig.~\ref{fig:states}(c)], $\Delta j > 0$ leads to decreases in the average applied current, while $\Delta j < 0$ leads to an increase in the average applied current. The modulated states are found adjacent to the IPP and OPP states, which suggests that variations in $\tau$ are not sufficient to destroy the self-synchronized oscillatory modes.

\begin{figure}
\centering\includegraphics[width=5.5cm]{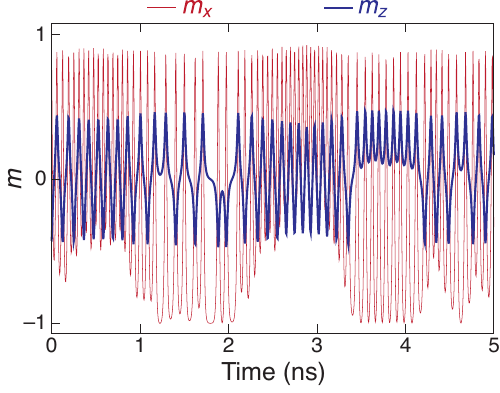}
\caption{Representative time trace of chaotic dynamics. $m_z(t)$ exhibits chaotic switching between the IPP and OPP modes.}
\label{fig:ipop}
\end{figure}

\begin{figure}
\centering\includegraphics[width=6cm]{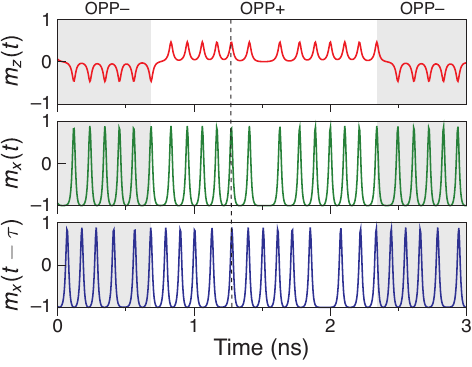}
\caption{Comparison of the time traces of the output and feedback signals in the chaotic regime. There is a synchronization between output ($m_x(t)$) and feedback ($m_x(t-\tau)$) signals before every mode switching (straight line) but there are also some synchronization events not followed by a mode switching (dashed line).}
\label{fig:synchrograph}
\end{figure}

When the time delay slightly exceeds the integer multiples of the precession period, signatures of chaotic dynamics appear. The dynamics largely comprises intermittent switching between the IPP and degenerate OPP states, with no well-defined periodicity. An example of the time dependence in this regime is shown in Fig.~\ref{fig:ipop}.
%%
%% Figure
In order to gain a better understanding of this chaotic regime, we examine the magnetization trajectories and feedback signals at the points where switching between the IPP and OPP modes take place. This is shown in Fig.~\ref{fig:synchrograph}, where $m_z(t)$, $m_x(t)$, and $m_x(t-\tau)$  are illustrated over several periods for $\tau$ = 0.5 ns and $\Delta j = -1$.
%%
%% Figure
Mode switching almost always occurs after a temporary synchronization between the output and feedback signals, as indicated by the solid lines in the figure. The second highlighted synchronization (dashed line) on Fig.~\ref{fig:synchrograph} is not followed by a OPP$+$ to OPP$-$ or OPP to IPP transition, but rather an extended dwell time in the OPP$+$ phase. As such, what appears to be a mode transition from the OPP$+$ to either the IPP or OPP$-$ state turns out to be a transient dynamics that brings the system back into the OPP$+$ state. It is therefore possible to have OPP$+$/OPP$+$ and OPP$-$/OPP$-$ transitions where a small transient phase occurs in between these states. This results in a jitter in the precession period, which may also impede subsequent synchronizations to the feedback signal.

\begin{figure}
\centering\includegraphics[width=6cm]{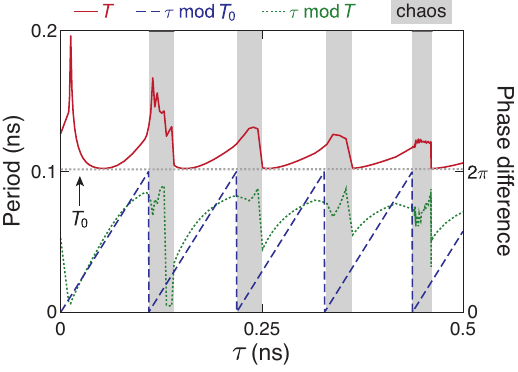}
\caption{Average precession period, $T$, and the phase difference between the oscillator output and feedback signal, as a function of the time delay $\tau$. Chaos arises when the delay falls in a small interval exceeding the quantity $\tau \, \mathrm{mod} \, T_0$, as indicated by the filled bands.}
\label{fig:finalphasediff}
\end{figure}

Since non chaotic behavior implies a fixed phase difference between the output and feedback signals (in the form of the delay), and mode switching is triggered by the synchronization of these two signals, it is interesting to examine how the phase difference between these two signals vary with the time delay. This is presented in Fig.~\ref{fig:finalphasediff}, where the oscillator period, $T$, and phase difference with the feedback signal, is shown as a function of $\tau$.
%%
%% Figure
$T_0$ denotes the precession period in the absence of chaos at $\Delta j = -0.1$. We note that other feedback strengths lead to the similar behavior and that certain aspects are analogous to the response to an ac current at fixed frequency.~\cite{Zhou:2007ke} The figure shows that the oscillator period exhibits large variations as a function of the delay, where the period almost doubles at small delays with deviations from the natural period decreasing with increasing delay. The appearance of the chaotic regime is intimately related to the phase difference between the feedback signal and the oscillator state. Consider first what happens when the IPP and OPP modes are attained. Here, the phase difference between the oscillator and feedback signals remain constant at a value $\tau \, \mathrm{mod} \, T$, where $T$ is close to $T_0$. Values of $\tau$ around a multiple of the natural period $T_0$ would therefore lead to a very small phase difference. However, Fig.~\ref{fig:synchrograph} shows that temporary synchronization leads either to mode switching or a jitter in the period. For the former, the system does not attain a stable limit cycle, while for the latter the jitter results in increases in the average period until the stable limit cycle is reached. These two cases are illustrated in Fig.~\ref{fig:finalphasediff}. For values of $\tau$ just below a multiple of $T_0$ (i.e., small negative phase differences), increases in the average period lead to stable oscillations, while for small positive phase differences a chaotic regime is attained. This occurs because mode switching takes place only at certain points along the trajectory, similarly to periodic core reversal in nanocontact vortex oscillators,~\cite{PetitWatelot:2012be} so chaotic dynamics can only appear if the feedback signal produces such transitions at certain points along the trajectories.

\begin{figure}
\includegraphics[width=5.5cm]{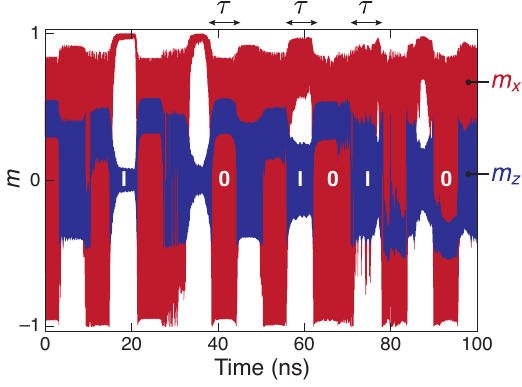}
\caption{Representative time trace of intermittence. The oscillator switches between IPP ($I$) and OPP ($O$) modes with a period that is close to the delay $\tau$.}
\label{fig:damped}
\end{figure}

Intermittency occurs for long time delays where $\tau \gg T_0$. As discussed above, this represents chaotic switching between well-defined IPP and OPP states. Such delays are comparable to the typical relaxation time toward the steady state orbit, i.e., the time required for initial transients associated with stable precession states like IPP or OPP to die out. In this regime, the oscillator settles into IPP or OPP states but switches intermittently between the two as in the chaotic state. An example of the time evolution is shown in Fig.~\ref{fig:damped}. 
%%
%% Figure
The time trace shows that the feedback drives near-periodic switching between the IPP and OPP states. After each switching event, the oscillator relaxes toward a stable oscillatory state, but transients that reappear in the feedback signal after a long delay causes the system to switch to the other oscillation state. Similar transitions are also observed between the IPP state and the static state where no oscillations are present. This is similar to the `oscillator death' scenario in systems of coupled limit-cycle oscillators.~\cite{Matthews:1990co} This behavior follows on from the different values of $\langle m_x \rangle$ attainable in the IPP phase [Fig.~\ref{fig:states}(c)], where $\langle m_x \rangle > 0$ combined with large $\Delta j < 0$ results in a suppression of the IPP mode and stabilization in the non-oscillatory state.

In summary, delayed feedback in a macrospin spin-torque nano-oscillator can result in a variety of dynamical states, where transitions between different oscillation modes can be triggered. The results suggest that delayed feedback may be a practical way for generating chaos and complex transient states in such oscillators, which might be useful for tasks such as fast random number generation~\cite{Uchida:2008gx, Li:2013il, Virte:2014bw}, chaos multiplexing for cryptography,~\cite{Scholl:1999} and chaos-based computing.~\cite{Ditto:2008gm}

J.K. acknowledges fruitful discussions with J. P{\'e}ter. A.A. acknowledges support from Conselho Nacional de Desenvolvimento Cient{\'i}fico e Tecnol{\'o}gico (CNPq, Brazil). This work was supported by the Agence Nationale de la Recherche (France) under contract nos. ANR-14-CE26-0021 (MEMOS) and ANR-17-CE24-0008 (CHIPMuNCS). The Chaire Photonique is funded by the European Union (FEDER), Ministry of Higher Education and Research (FNADT), Moselle Department, Grand Est Region, Metz Metropole, AIRBUS-GDI Simulation, CentraleSup{\'e}lec, and Fondation Sup{\'e}lec.

\bibliography{articles}

\end{document}